\documentclass[12pt,preprint]{aastex}

\begin{document}

\title{A Realistic Cosmological Model Based on Observations and
Some Theory Developed over the Last 90 Years}

\author{G. Burbidge, University of California San Diego, La Jolla, CA 92093-0424}

\begin{abstract}

\underline{Introduction}

This meeting is entitled ''A Century of Cosmology.''  But most of the papers being given
here are based on work done very recently and there is really no attempt being made to
critically review what has taken place in the last 90 or 100 years Instead, in general
the participants accept without question that cosmology equates to ''hot big bang
cosmology'' with all of its bells and whistles.  All of the theory and the results
obtained from observations are interpreted on the assumption that this extremely popular
model is the correct one, and observers feel that they have to interpret its results in
terms of what this theory allows.  No one is attempting to seriously test the model with
a view to accepting it or ruling it out.  They are aware, as are the theorists, that
there are enough free parameters available to fix up almost any model of the type.

The current scheme given in detail for example by Spergel et al (2006, 2007) demonstrates
this.  How we got to this stage is never discussed, and little or no attention is paid to
the observations obtained since the 1960s on activity in the centers of galaxies and what
they imply.  We shall show that they are an integral part of a realistic cosmological
model.

In this paper I shall take a different approach, showing first how cosmological ideas
have developed over the last 90 years and where mistakes have been made.  I shall
conclude with a realistic model in which all of the observational material is included,
and compare it with the popular model.  Not surprisingly I shall show that there remain
many unsolved problems, and previously unexpected observations, most of which are ignored
or neglected by current observers and theorists, who believe that the hot big bang model
must be correct.
\end{abstract}

\underline{The Beginning: 1915-1936}

The major starting point was Einstein's development of general relativity.  In applying
his gravitation theory to the universe as it was then know (the Milky Way) he tried to
explain a static universe.  He could only do this by introducing the cosmical constant.
This was Mistake 1 made by a great theorist who was misled, because no one realized how
large and complex the universe was, and the astronomical community barring a few notable
exceptions like Heber Curtis all believed that everything was contained within the static
Milky Way.

In 1922 and 1927 Friedmann and Lemaitre independently showed that Einstein's equations
allow both expanding and contracting solutions.  This work came at about the time Hubble
and others using the spectra of spiral nebulae obtained by Besto Slipher at the Flagstaff
Observatory showed that there was a good correlation between the redshifts and the
apparent magnitudes of the nebulae (the fainter nebulae, the larger were the redshifts).
It already had been shown by Hubble in 1924 that the spiral nebulae (galaxies) lie
outside our Milky Way and are independent ''island universes''.

Overwhelmingly, the majority concluded that the redshifts must be due to expansion.  This
could have been wrong, as was suggested by Fritz Zwicky, Edwin MacMillan, Max Born, and
others, but this view (the tired light hypothesis) goes against atomic physics and is
generally neglected.

Since the observed expansion and the Friedmann solutions to Einstein's equations came at
about the same time, by about 1930, it was generally accepted that the universe must be
expanding.  Thus, if energy is conserved, this means that extrapolation back in time
forces us to the conclusion that early in its history, the universe was very small and
compact.

At that time nuclear physics was in its infancy so that no realistic discussion of the
nuclear physics of an early universe could be made.  This is seen from the work in 1936
by Lemaitre in his paper entitled, ''The Primeval Atom,'' and also by studies by
Houtermans and Atkinson.  However, by the late 1930s experimental nuclear physics
involving the lightest isotopes had seen dramatic progress, and by 1938 Hans Bethe, C.F.
Weitzacker, Charles Crichfield and others had shown that at high temperatures and
densities in the centers of stars, neutrons, protons, electrons and positrons, neutrinos,
and radiation will interact through a series of reactions so hydrogen can be burned to
produce helium, the next most abundant nucleus to hydrogen in the periodic table.  Thus
it was possible that helium could be produced in an early universe.  But in thinking
along those lines George Gamow and others were already unconsciously rejecting another
possibility.  This was the possibility that while the universe is expanding, it always
remains the same.  If this was correct, the universe never had a young, dense phase.

Nearly twenty more years were to pass before T. Gold seriously made this proposal in
1947-1948.  He made it to Hermann Bondi and Fred Hoyle.  They worked out the consequences
of this idea, and the three of them published two papers on the steady-state cosmology
(Bondi {\&} Gold, 1948; Hoyle 1948).  This turned out to be a very unpopular theory.  it
requires that creation takes place not in an instant at a beginning, but throughout the
cosmos, at a steady rate determined by the rate of expansion (i.e. in modern terms it
requires dark energy).

In this case the universe will not decelerate, as must be the case for the Friedmann
models, but as energy is added it can maintain a steady expansion or even accelerate.  In
field theoretical terms this theory requires the existence of negative energy fields,
which in the 1940's was the kind of physics not considered acceptable by the theoretical
physics establishment.  This is one of the reasons why the Hoyle-Narlikar C field theory
of the 1960s (Hoyle {\&} Narlikar, 1964) was so widely ignored.  But now following the
reports that the universe is accelerating, this type of creation (the so-called dark
energy) has found acceptance.  Today's phantom fields are recognized to be the C field of
Hoyle and Narlikar in a different garb.

The neglect of this alternative, and the hostile approach taken to those who proposed the
steady-state model was \underline{Mistake 2}.  However, it is fair to point out that
while many of the observational objections to the steady-state turned out to be wrong,
(cf Hoyle 1969), ultimately it was shown that the universe is evolving.  Thus the simple
steady-state model can be ruled out.

\underline{The 1940's}

The next major step came with attempts to understand the origin of the chemical elements.
Measurements of the relative abundances of the elements in the periodic table in
meteorites and in the sun suggested that some of the gross features of the abundance
curve were related to the nuclear structure of the nuclei. But first it was important to
understand where the most abundant element after hydrogen - namely helium - had been
made.

In the late 1940's many leading physicists, including George Gamow, Edward Teller, Enrico
Fermi, Maria Mayer, Rudolph Peierls, and others, took the position that the most likely
place of origin of the elements must be a place where there is likely to be a large
baryon density, out of which the heavier elements could be built up from protons,
neutrons, electrons, positrons, neutrinos, and radiation.  The speculated that this must
have been the early universe.  While it was known that helium was being synthesized in
stars, there was far too much of it for it to have been simply been built up in stars
over the age of the universe.  If we live in a Friedmann universe the time available is
{$\approx(H_{o})^{-1}$ where $H_{o}$ is the Hubble constant.  At that time Hubble and
Humason had given a value of $H_{o} of 550 km sec^{-1}Mpc^{-1}$, corresponding to an age
of only about $2 x 10^{9}$ years.  This was nothing like enough time, though Gamow and
his colleagues did not know that over the next 40 years the value of $H_{o}$ would be
decreased by a factor $\approx10$ giving an age $\approx15$ Gyr.  Even then this does not
give enough time for helium production.

But in order to make nucleosynthesis in an early universe work at all, George Gamow and
Alpher and Herman (cf Gamow 1946; Alpher {\&} Herman, 1950) found that the had to
\underline{choose} a value for the initial baryon/photon ratio in the ''beginning,'' a
value radically different from what had been assumed before.  It turns out that photons
must dominated over baryons by a very large factor $\approx10^{11}$ in such an initial
synthesis scenario.  Earlier it had been assumed that the baryons dominated over the
photons.  (The value that Gamow et al chose, is very close to the value used today}.

Even when this was done, it was shown that only $^{2}D, ^{3}He, ^{4}He and some ^{7}Li$
could be built, since there is no stable mass 5 or 8.  Having found this, most of the
people working in this area left it, since it was clear that even if there was a hot big
bang, only the most abundant element, helium, and the very light isotopes could have been
made in it.

However, by then Gamow, Alpher and Herman and others were really attracted to the idea of
a beginning, and a hot fireball of radiation, even though there was no proper theoretical
basis for believing in it.  Gamow et al predicted that as it expanded the hot fireball
would maintain its Planckian form.  Thus they predicted that if the radiation from the
hot fireball was found  it would have a black body form.

If such a situation ever did exist it is possible to calculate in detail the relative
abundances of the light isotopes, particularly deuterium, and compare the results with
observation.

In modern times there have been detailed calculations of the relative abundances of the
light isotopes which will be made, and they have been compared with measured abundances,
particularly of $^{2}D$ which can be detected in the spectra of QSOs with high redshifts.
It is very difficult to make such measurements and it is a tour-de-force to achieve this,
as Tytler et al (2006) have done better than almost any one else.  It is often claimed
therefore, that the abundances calculated originally by Gamow and his collaborators, and
later by many others agree so well with the observed abundances, that this is proof that
the big bang occurred.  But this is simply not correct.  The statement that the big bang
theory \underline{explains} the observed microwave background and also
\underline{explains} the helium abundances is to distort the meaning of words.

Explanations in science are normally to be considered like theorems in mathematics, to
flow deductively from axioms and not to be restatements of axioms themselves.

Thus the radiation-dominated early universe is an axiom of big bang cosmology, and the
supposed explanation of the  CMB,  and the light element abundances, is a restatement of
that axiom.  To reiterate, the baryon density and temperature relation has to be fixed
suitably in order to explain the light element abundances.

Thus it must be remembered that the whole argument is based on the idea that helium
\underline{was} made by such a fireball, and much as most people want to believe it there
is no independent evidence that this ever did take place.  The blindness of the community
to this point is \underline{Mistake 3}.

What about the origin of all of the elements?  Most of them cannot have been made in a
big bang even if there was one.

The answer came from Fred Hoyle.  In 1946, having noticed that the observed abundance
curve showed a mild peak in the vicinity of iron - the region where nuclei are most
strongly bound where the packing fraction is largest, he concluded that the heavy
elements must have been made in some kind of equilibrium process when the mix must have
reached a temperature $\approx{3 x 10^{9}}$ (Hoyle 1946).  He concluded that this must
have taken place in stars.  Thus the theory of stellar nucleosynthesis was born.

Hoyle followed this up with a more detailed study in 1954 (Hoyle 1954), and this in turn
led to the extensive investigations of stellar nucleosynthesis by $B^{2}FH$ and Cameron
in 1957 in which the case was made and largely established that ll of the isotopes (with
the exception of $^{2}D, ^{3}He, ^{4}He, and ^{7}Li$ were made at different stages of
stellar evolution.  By then a great deal of observational evidence had been found showing
that individual stars could be observed to be the synthesizing element.  For example, the
discovery of $^{43}Tc$ by Merrill.  Clearly different properties of nucleosynthesis go on
at different stages in the evolution of stars.

\underline{The 1950's Onward}

What is clear from this is the very large amount of energy that must have been released
in the production of helium from hydrogen, and that this must still exist somewhere in
the universe.

In the 1940s cosmologists had no idea where such energy could be, but in retrospect we
can also see now, that the most likely scenario is that it is contained in the cosmic
microwave background.  Back in 1926 Eddington had tried to estimate the amount of energy
present in starlight in our galaxy, and in recent time this has been re-calculated by
Pecker and Narlikar (2007).

The first indirect measurements of the microwave background radiation came in the late
1930's following the detection of absorption by interstellar molecules due to CH, CN, and
$CH^{+}$ by Adams at Mount Wilson and McKellar at the Dominion Astrophysical Observatory.
From purely observational considerations McKellar (1941) showed that the temperature of
this radiation (if it is in black body form) lies in the range ${1.8^{o}K<T<3.46{o}K}$.
In view of the later work this was a remarkably accurate estimate.

In the 1950s and 1960s a new generation of physicists and astronomers attempted to
determine what is the correct cosmology.  Ever since the  1930s most physical scientists,
theologians, and philosophers have wanted to believe that we live in an evolving universe
which had a beginning.  In my view this all began because the discovery of the expansion
was made in the same period when the Friedmann-Lemaitre solutions of Einstein's equations
were found.

What actually happened?

In the late 1950's - early 1960's, Robert Dicke at Princeton became interested enough in
Gamow, Alpher and Herman's ideas to try to detect the radation which the hot big bang
believers had predicted would be present.  The temperature could not be predicted,
essentially because there is no basic theory for the big bang (cf Turner 1993), but it
was believed that if the radiation was detected and had black body form this would prove
that the big bang picture was correct.

Similarly the Russian cosmologists, led originally by Zeldovich believed passionately in
this idea.  This is clear from the fact that in Moscow, before the CMB was identified
observationally, the Russian school referred to it as ''relict radiation.''

From that time on one huge danger that has been present and that continues to be be
present is that all of the work on the CMB has been carried on by observers who are
absolutely convinced that whatever they find, they are quite sure where it came from.
This has led to a band wagon now so overwhelming that alternative interpretations of the
data are hardly ever mentioned, never taught, or discussed at meetings, or referred to in
textbooks.

What actually happened?  In 1965 Penzias and Wilson found the CMB (Penzias {\&} Wilson
1965), and by 1992 it had been shown to have a black body temperature of $2.728^{o}K$
(Mather et al 1990, 1994), right in the middle of the range given by McKellar (1941),
though he was never given any credit for this, and most cosmologists have never heard of
his name.

\underline{If} it was generated in a big bang, then as Gamow and his colleagues proposed,
it should have a black body spectrum.  And of course the results obtained in space, most
recently in the famous WMap observations, show the blackbody nature is well established.

But this is not in any sense proof that a big bang ever took place.  Provided that the
CMB is universal, and not contained to our own Milky Way, a big bang, \underline{or any
other process} which is able to provide the energy in black body form is acceptable.

There are two lines of argument which suggest that the CMB is universal.  It has been
shown that a very small temperature asymmetry in the CMB is most likely due to the fact
that our Milky Way is moving through the CMB with a velocity $\approx500 km sec^{-1}$.
Secondly, it was shown many years ago by Greisen, Zatzepin and Kusmin, that if the CMB is
universal, at the highest energies, around $10^{19.5}ev$, the cosmic ray proton spectrum
should be truncated, due to collisions between very high energy protons and photons of
the CMB.  This effect (the GKK effect)(a knee or at least a flattening in the spectrum)
has recently been detected.  Thus the truly extragalacatic nature of the CMB has been
established.

In $B^{2}FH$ we omitted any discussion of the origin of helium, though we were well aware
of the problem.  In 1955 Bondi, Gold, and Hoyle (1953) still believing in the simple
steady state cosmology, proposed that the energy must have been emitted from K giant
stars, and in 1958, not being aware of that work, I looked at possible places of origin
of the helium(Burbidge 1958) and concluded that, the time scale for production of helium
from hydrogen burning in stars was still too short, even if the universe had a beginning
some 10 billion years ago.

Thus I concluded either that it had been made in a big bang, or it may have been made in
short periods in the evolution of galaxies when they become much more luminous than they
are for most of their radiating lives.

But what both Bondi et al, and I, paid little attention to, was the significance of the
value of the energy density that must be present. Neither of us paid attention to the
fact that this value, $4.5 X 10^{-13} erg cm^{3}$, which is derived simply by using the
observed values of the mass density in galaxies and the observed He/H ratio, leads
directly to the conclusion that if all of the energy released in the conversion of
hydrogen to helium is still present, and has been degraded to black body form, the
temperature will be about $2.75^{o}K$, which we know now \underline{is} almost exactly
the temperature of the universal CMB.  In my view this cannot be a coincidence and is the
key to our understanding.  Overlooking this was \underline{Mistake 4}.

It strongly suggests that the CMB does arise from hydrogen burning, but the time scale
for the universe must be much longer than $(H_{o}^{-1})$.  Since the universe is largely
made up of condensed regions of matter (lumps) in the form of galaxies, these must be the
places where the creation processes occur.

This leads us directly to the idea that the CMB arises in active galaxies, and the
overall time scale leads us to the conclusion that the universe is cyclic.  This means
that we are in an expanding phase now, with a cycle time of $\approx20$ Gyrs.  Later on
the universe will slow down and start to collapse.  However, the pressure exerted by the
active galaxies as they squeeze close together means that the universe will not collapse
back down to a region of extreme conditions.

By the time the galaxies begin to overlap the activity in the centers will generate
enough energy and pressure so that the collapse will slow down, and then begin to expand
again, thus giving rise to a new cycle.  The overall timescale must be $>10^{12}$ years.

Since the 1960's it has become clear that the nuclei of galaxies can be extremely active.
Thus it is natural to believe that they make up the discrete sources where little big
bangs occur\footnote{It is of some interest to point out that all of the other less
energetic diffuse radiation fields-radio, infrared, optical and X-ray-are believed to
arise from large numbers of discrete sources.}  The realistic model that this chain of
evidence leads to drives us to the conclusion that creation must be taking place, but it
occurs in the nuclear regions of galaxies, near massive centers which my be black holes.
It is at the level that ''new'' physics, almost certainly involving a breakdown of
Einstein's theory in the strong field limit, is required.

The model does not require that there was a unique beginning, i.e. no big bang.

The scenario that so many believe in today with a single beginning requires a series of ideas
for which there is little or no, evidence.  Those who believe in the model are driven by the
over-riding (correct) belief that without a series of extra hypotheses that their model won't
work.  The hypotheses that are required, include an inflationary phase, (for which there is no
independent evidence), initial density fluctuations (some believe of quantum origin, but no one
knows), a non-baryonic matter which dominates over normal matter (for which there is no
evidence), but without which galaxies cannot be formed, an absolute belief that all close
interacting galaxies are coming together and never coming apart, etc.  With all of these
assumptions in place, strong claims are made that together with the CMB everything can be
understood.  But their model required that less than 5% of the mass energy in the universe
is baryonic!

\underline{Active Galaxies as the Most Likely Sites of Creation}

\underline{Energetics}

It was the discovery that powerful non-thermal radio sources often originate in apparently
normal elliptical galaxies, and that some spiral galaxies show evidence of non-thermal optical
energy in their nuclei (Seyfert galaxies) that led to the idea that centers of galaxies can
literally explode and led to the idea of active galactic nuclei (AGN); an early review was
published in 1963 by Burbidge, Burbidge and Sandage (1963).

Calculations of the amount of energy seen as a result of these nuclear events led to the
conclusion that non-thermal energy in the range $10^{55}-10^{60}$ ergs or
$\approx10-10^{6}M{\odot}c^{2}$ was released in times $\approx10^{6}-10^{8}$ years.

It was already clear in the 1960s that energies of this magnitude could, in principle,
arise only from gravitational energy release due to matter falling on to a massive
central black hole, or they must be due to the creation process (Hoyle, Fowler, Burbidge
{\&} Burbidge, 1964).  From the earliest days, the community embraced the view that the
gravitational collapse mechanism must be the correct one.  They relied heavily on leading
theorists, particularly Martin Rees and Roger Blandford who made elegant models to
explain these events - particularly the radio sources, using conventional physics.

This approach was to be expected since the release of gravitational energy is a process
based on known physical theory, whereas invoking creation for which there is no
well-developed theory is not a view that most scientists wish to take.  This follows the
prevailing view, that observed astronomical phenomena must be explained using known
physics, and never the other way around, that from observations we can learn new physics.
This despite the fact that observations often have led to the development of new physical
theory, as was the case with Newton and others who succeeded him.  I believe that her
another major mistake (Mistake Number 5) is being made.

In this case my reasons are as follows:  Unless the distance scale for the galaxies is
completely wrong, we cannot escape from the view that what we see in the active objects
is the generation of very large fluxes of relativistic particles and/or very large fluxes
of photons with energies ranging all the way from radio to X-ray frequencies with a
spectrum with a non-thermal index. In some cases the photon flux is seen to be variable
in time meaning that the largest volume out of which the energy is generated must be as
small as $10^{15}-10^{16}cm$, and in the largest sources--the powerful radio galaxies,
the sizes can range up to $\approx10^{5} kpc$, or $10^{23}cm$.

But if the ultimate energy source is associated with a massive black hole, classical
theory tells us that energy can only be released very close to the Schwarzchild radius,
as gravitational radiation, and the maximum efficiency obtained for a rotating black hole
(the Kerr solution) is only about $8\%$.  This energy has to be transformed and got out
to very much larger regions where it can be detected either as photons, rapidly moving
ejected gas, or large fluxes of relativistic particles which give rise to synchrotron
radiation. The processes of energy transfer to different forms, and to much larger
regions, are bound to be highly inefficient.

This view was brought home to me more than fifty years ago at the Solvy conference of
1956 where I described the very large energetics of the radio galaxies.  The leading
physicists immediately asked me whether I had any idea of the efficiency with which we
can produce beams of relativistic particles on the Earth, i.e. in accelerators.  With aid
of Emilio Segre, one of my interrogators, I found that the efficiency of a linear
accelerator (at Stanford) was about $1\%$ and for a proton synchrotron, the efficiency is
about $0.1-0.01\%$, i.e. the ratio of energy out (beam) to energy in (the power station)
lies in the range $1-^{-2}-10^{-4}$. Independently Richard Feynman also pointed out to me
that in making a proper calculation of the efficiency of such a machine (he liked to call
M87 the M87 Synchrotron) account must be taken of the energy required to
\underline{build} the machine.

I have wrestled with this problem of efficiency ever since.  I don't have a solution, but
I believe that since the efficiency must be very low, the common view that we are just
seeing matter falling into a massive black hole is fallacious.

If it were true, not only must the masses in the centers of galaxies be many orders of
magnitudes greater than any detected so far, but where is all of the energy which is not
seen?  If the efficiency is $10^{-3}$ where is the $999/1000$ of the energy which was
released? This view is supported to a considerable extent by the detailed studies of the
physics of the jet in M87 (Sunyaev et al 2005) which contains a massive black hole
$\approx10^{8}M{\odot}$ and where the efficiency calculated from the non-thermal jet
emission $\approx10^{-3}$.

We do have direct evidence from dynamical studies that many nearby quiescent galaxies
have central dark masses (black holes) with masses $0^{6}-10^{7}M{\odot}$.  For the
majority of galaxies we cannot get a direct measurement of the mass in the nuclear
region.  However, a correlation has been found between the luminosity (assumed to be
concentrated in the nucleus) and the masses of the nuclei of a few nearby AGN.  Here the
mass has been determined using the virial theorem using velocities of the rapidly moving
gas (it is suspect in many cases since there is no evidence that the virial holds).

However using the correlation between luminosity and mass allows astronomers to estimate
masses of high redshift systems.  In all of these estimates they assume an efficiency of
conversion of gravitational energy to luminosity $\approx10\%$-very close to the Kerr
limit.

We are then told that very distant, very luminous galaxies have masses of
$10^{12}-10^{13}M{\odot}$.  But of course, if the redshifts of these objects are measures
of distance, and to a very much lower efficiency is accepted, the masses would have to be
$10^{14}-10^{15}M{\odot}$ and any conventional scheme for making galaxies fails.

Thus I believe that creation in the galactic nuclei in situations in which Planck
particles are produced and the Hoyle-Narlikar creation process is at work is much more
likely.  In the creation process the efficiency must be at least an order of magnitude
greater than that which can be properly be assumed in the conventional approach.

\underline{Quasistellar Objects}

The discovery of active galactic nuclei (AGN) in the 1960s in the hands of the radio
astronomers was immediately followed by the discovery of the radio emitting quasi-stellar
objects (QSOs), and Sandage and others showed that QSOs which were not radio emitters
were frequently present.  Thus as Sandage somewhat rashly put it, a new constituent of
the universe had been found.

How are these objects related to galaxies, and particularly to AGN, the distributed
centers of creation as I have just described them?

The major property of QSOs which was totally unexpected is that they are highly compact
and have the appearance of stars in our own galaxy, but have very large redshifts, as was
first shown by Schmidt and Oke.

It was this latter property that led the community very early on to conclude that they
must be very distant and very luminous.  Thus it was assumed that they must be giving us
information about the earlier state of the universe, and this meant that they could be
used for cosmological investigation.  But from the earliest days their properties were
hard to understand in this way.  Their apparent brightnesses are not smoothly correlated
with their redshifts as is the case for normal  galaxies.  Thus if the redshifts are of
cosmological origin there is a very wide scatter in the bolometric luminosities.  Since
the optical radiation is non-thermal this might well be the case.

Also the optical spectra of QSOs are dominated by strong, broad emission lines due to
rapidly moving hot gas and thus they are very similar to the spectra of the nuclei of the
class of AGN known as Seyfert nuclei (Seyfert 1943).  Thus most astronomers immediately
subscribed to the continuity argument first published by Krirtian (1965) who argued that
at low redshifts it is easy to see the whole galaxy with its Seyfert nucleus, but as we
study fainter and more distant objects with intrinsically brighter nuclei and larger
redshifts the stars outside the nucleus are harder to detect and we only see the nucleus
which we call a QSO.

When Kristian first showed the correlation based on very few objects it looked very good
and this interpretation was widely accepted, even though he excluded from his plt the
brightest QSO, 3C 273 which of course did not fit the correlation.

Later on many low redshift QSOs with $z\leq0.2$ were observed with the Hubble telescope
by Bahcall et al and were shown to lie in genuine spiral galaxies.  There are also some
bare QSOs. But the general problem remained.  When low luminosity ''fuzz'' is seen around
the image of a QSO is it always starlight, or is it high temperature gas with the same
redshift?  If it really is starlight and the continuity argument is correct, a spectrum
should always show the normal stellar absorption spectrum with the same redshift as the
emission-line QSO.  In a number of the first cases studied for some of the bright QSOs,
it was shown that the fuzz was not due to stars, but due to gas with narrow emission
lines at the same redshift as the QSO.  The fuzz was not starlight.

In only one of the early QSOs, 3C 48, was it shown that a galaxy was present at the same
redshift.  In this case the QSO is \underline{not} symmetrically placed with respect to
the QSO.  As time has gone on there have been further elaborate studies of comparatively
low redshift systems, and the view has grown up that the high redshift QSOs must all lie
in elliptical galaxies.  The only unambiguous way to test this is to study a sample in
which the emission line redshift is \underline{not} \underline{known} and the redshift of
the fuzz is determined first.

What is actually done is to do surface photometry; of the faint halo (fuzz) around to the
QSO and then compare this with what you would expect to find in a giant elliptical
galaxy.  We still have no absorption line spectra of the fuzz around high redshift QSOs
in which it can be shown that the two redshifts agree.

And this is one field in which no further progress will be made since everyone already
\underline{believes} that all QSOs lie in galaxies at the same redshift (it's in the
textbooks, and you wouldn't get observing time to look, and after all for most people,
AGN = QSO = AGN, by definition).

The problem gets even worse when we come to the redshifts.  In the late 1960s several
more observed properties of QSOs were found. First it was found that the optical and the
radio flux was variable in time.  This immediately sets a limit on the size of the
emitting region. A more difficult problem immediately arose when it was realized that the
non-thermal synchrotron radiation model would be self-quenching because the process would
generate a photon energy density equal to the magnetic energy density.  There are only
two ways out of this.  One is to argue that very large coherent relativistic motions are
present so that very large values of $\gamma=(1 - v^{2}/c^{2})^{-1}$ are involved.

The second is to argue that the variable sources are not the distances given by their
redshifts - they are much closer, so the energy densities in the sources are much lower,
and the contradiction is removed.

In 1966 when we brought up this dilemma (Hoyle, Burbidge, and Sargent 1966) Martin Rees
(1966) and L. Woltjer (1966) immediately opted for and led the community to believe that
highly relativistic motions were the answer.  This was not surprising since the
alternative explanation would mean that we had no explanation for the redshifts, and QSOs
could not be used for cosmology.

Soon after this rapid angular motions began to be detected interferometrically in radio
sources with both low and high redshifts.  By now such motions have been found in many
sources and we are told that for all high redshift objects superluminal motions are
present, and they require large values of $\gamma$.  However I have shown that in many of
the sources there is a low redshift galaxy close to, or part of the high redshift
variable radio source, and if this is used to determine the distance of the object rather
than the redshifts of the non-thermal sources the velocities are high but $\leq c$
(Burbidge, 2004) not highly relativistic i.e. the superluminal motions are an artifact of
the redshift.

The most difficult problem involving the QSOs was brought to light by Arp in 1987 (see
Arp 1987) and others (Burbidge, Burbidge, Solomon and Strittmatter, 1971; Arp, H.C., et
al, 2002) who found statistical evidence suggesting that many high redshift QSOs lie so
close to low redshift galaxies that they must be physically associated, so the QSO
redshifts are \underline{not} due to the expanding universe.

Arp's results are spectacular and are well known.  His evidence and his persistence in
putting it forward led his professional colleagues at Mount Wilson and Palomar in the
early 1980's to having him removed from the telescopes, and he has lived and worked at
the Max Planck Institute in Munich for the last 25 years.

This is one of the worst cases of discrimination involving scientific goals that I have
ever seen.  But it worked.  The younger generation saw what happens if you don't follow
the party line -- no observing time, no staff or faculty positions, no financial support,
no conference invitations.  But to go back to redshifts.

For the last 25 years or more a small group involving particularly H.C. Arp, the
Burbidges, the late Fred Hoyle, William Napier, and J.V. Narlikar, and a few others have
shown that there is much observational evidence which shows that many bright QSOs with
high redshifts are associated with low redshift, nearby galaxies many of which are highly
active.  One of the most recent and most spectacular discoveries was made by P.
Gallianni, Margaret Burbidge, and others who found a QSO with a redshift $z = 2.11$ only
$10\prime\prime$ from the center of a nearby galaxy NGC 7619 with $z = 0.001$ (Gallianni
et al 2005).  Since only a small number of the associations between high and low redshift
objects show a luminous bridge e.g. (NGC 4319 and MK 205), (Arp, 1987); NGC 7503 and its
companions, (Lopez-Corredoira {\&} Gutierrez, 2007) the majority of the associations can
only be tested by statistical techniques.  Thus the results are always treated as
controversial, largely because the referees of the papers and sometimes the editors don't
want to believe the results.  Despite prejudice quite a lot has been published, but then
ignored.

By now there is general agreement that the observed effect \underline{is} real, since
statistical evidence of this clustering has been found even by groups who believe that
all redshifts must be cosmological in origin.  They try to argue that the effect is due
to gravitational lensing by large amounts of dark matter which magnify the faint
background QSOs lying close to the lower redshift galaxies, an ingenious argument first
put forward by Canizares (1990).  Of course the models don't really work, and as we shall
discuss later there is no independent evidence for the widespread existence of
non-baryonic dark matter.

\underline{Dark Matter}

The theory of stellar structure is one of the best understood branches of modern
astrophysics and it is well established that dark matter is a natural end product of
stellar evolution. Also stars with very small masses may be too faint to detect.  Thus we
expect that some dark matter will be present in the form of brown dwarfs, white dwarfs,
neutron stars and black holes.  Also diffuse matter in various forms may not be
detectable through the emission or absorption of radiation.

Thus we can expect that dark matter is likely to be present in galaxies.  Consequently it
was not surprising when radio techniques and optical techniques led to the detection of
the so-called flat rotation curves of spiral galaxies.  These have been generally
interpreted as meaning that in outer parts of flattened systems and in their halos there
is much matter that cannot be seen but is exerting gravitational force.

Estimates of the amount of this dark matter have led to the idea that the total masses
are much greater than those estimated from the visible stars, gas and dust, often by an
order of magnitude or more.

The more general used of the virial theorem where it is \underline{assumed} that all
systems involving many galaxies are in equilibrium, leads to the general conclusion that
there must be large amounts of dark matter in or between the galaxies in groups and
clusters.  In the hands of theoreticians this had led to the general belief that the
larger the physical system is, the greater is the mass and the fraction of dark matter.
This idea has been extended all the way from pairs of galaxies up to rich clusters.  But
the correlation has only been established by supposing that all of these systems are
bound, i.e. the argument is circular.  It has been supported by the discovery of large
amounts of hot gas emitting X-rays in some rich clusters, and by the claim that evidence
from gravitational lensing due to the presence of foreground clusters leads to the
conclusion that large amount of unseen mass are always present.  In support of this there
clearly are many clusters where there is every indication that they are smooth and
relaxed so that the virial argument is likely to be correct.  But as is often the case in
the band wagon approach it is now assumed that this is a general property of groups and
clusters.  Observationally it is true that in almost every case that the kinetic energy
of the luminous systems appears to be much greater than the potential energy due to
gravitational interaction.  But this argument is taken to extreme lengths when pairs, or
small interacting groups of galaxies are considered.  For reasons not apparent to anyone
outside the field, if two galaxies are seen to be interacting together it is always
assumed that they are \underline{coming} \underline{together} or are in equilibrium.  It
is never assumed that they may be \underline{coming} \underline{apart}.  In the best
cases it is possible to decide which, since if they are falling together tidal tails may
be detectable.  But in most cases, one cannot tell.

Why is the argument made this way?  Because if they are coming together this is
compatible with gravitational theory, but if they are coming apart we don't know why.
Also in the standard galaxy formation scheme required in the big bang scenario, small
galaxies are aggregated together to form more massive systems.  In other words, while we
might learn something new if we seriously considered that observations were telling us
that in some cases galaxies are coming apart, for example perhaps galaxies beget
galaxies, this possibility is ignored.  Apparently because we don't want to seriously
consider anything new.

In the 1960s Victor Ambartsumian studied very carefully many observations of clusters and
groups, and made the radical suggestion that since many of the systems do \underline{not}
appear to be in equilibrium, perhaps they are not in equilibrium, and they may be coming
apart.  His arguments were ignored, for two reasons, both based on the belief that the
hot big bang cosmology is correct.  The first reason, as it was made by Jan Oort, is that
if groups of galaxies are coming apart, we don't know where they are coming from, and
since their lifetimes in the groups are short, they cannot have all been made in the
early universe.  The second simply as was stated above if they are exploding apart, we
don't know why.

Neither of these objections is scientifically based, but as is the case in other aspects
of cosmology, the fix is in, and belief in the use of the virial is overwhelming.

A second approach concerning the interactions between galaxies is the suggestion made by
Milgrom (1983) that Newton's law needs to be modified for large distances. This idea was
more recently put on a more firm theoretical basis by Beckenstein (2004).

With the use of this modification of Newton's law, it is possible to explain the flat
rotation curves of spiral galaxies and some other phenomena, without requiring the
presence of dark matter.

All of the observational evidence described so far is concerned with dark matter, and
until theoretical cosmologists got into the act it was assumed without question that dark
matter is made up of baryons, because this is all that we have direct scientific evidence
for.

However, one of the major problems of cosmology is to understand the origin of galaxies.
If the conventional big bang scheme is followed, the galaxies must have been formed early
in the universe by the growth of density fluctuations (assumed to be present initially)
in the hot expanding cloud.  McCrea and others more than 60 years ago showed that such
growth will not occur in a hot expanding cloud.

This could well be seen as a real objection to this cosmological model, but the general
view that has been developed is that since the model must be correct, the galaxies must
have been formed.  For this to occur it has to be \underline{assumed} that there must be
another component of mass present which can exert gravitational force but nothing else.

This has now been inserted into the theory and is dignified by the title ''non- baryonic
dark matter'' (WIMPS).  With this ingredient which must dominate, it is theoretically
possible to simulate galaxy formation.  Thus beautiful simulations of galaxy formation
have been made by Simon White and Mike Norman and their groups, and others.

They make comparisons of their models which are all of dark galaxies, with what we see
observationally in light, and it is claimed that the large scale structure seen in the
visible galaxies is compatible with the models.

A second purely theoretical argument for the presence of non-baryonic dark matter is also
completely based on the belief that there was a hot big bang.  This comes from the fact
that the amount of ordinary matter seen falls far short of the mass that must be present
if primordial nucleosynthesis took place (Fukugita, Hogan, and Peebles 1998). Thus the
shortfall is assumed to be made up of non-baryonic matter.

While many observations are now being directed towards the detection of dark matter, even
if any of them is successful it will not be able to distinguish normal matter (which we
know exits) from non-baryonic matter, for which there is not observational evidence at
all.

Dark matter is certainly required in both the big bang and in the cyclic universe models.
But in the cyclic universe model there is no reason to believe that anything other than
normal matter is required to be present.

However, without the presence of non-baryonic matter the conventional cosmology fails.

If MOND turns out to be correct, and/or if in many situations the virial does not hold,
the case for a very large amount of dark matter is severely weakened.

But the state of mind of the community is such that much effort is being exerted (and
much money is spent) on looking for dark matter on the assumption that it must be there.
Most of the observers no longer bother to distinguish between baryonic dark matter and
non-baryonic dark matter, though one is real, and one may not be.  And of course no one
dares to make this argument to the funding agencies.

\underline{Acceleration in the Universe}

Nearly 10 years ago, using supernovae of Type Ia as standard candles, Perlmutter, Riess
et al (1999) reported that at redshifts in the range 0.5 to 1 the curvature in the Hubble
relation suggested that the expansion is accelerating.  Several reports since then have
supported these initial results (cf Riess et al).  There are still a few very experienced
observers, particularly Allan Sandage, who still doubt the reality of the effect.  but if
real acceleration is present, this agrees with the prediction made for the classical
steady state model (cf Hoyle and Sandage 1956).  It shows directly that matter creation
is taking  place long after any initial explosion may have occurred.  However, the
physicists who discovered this effect either were completely ignorant of the predictions
of the past, or chose to ignore them.  Thus they never mentioned the possibility, but
instead the incorporated the results into the standard Friedmann model which they believe
in, by bringing back to life the cosmical constant, which now claimed to be positive,
(but according to the latest results has to vary with epoch) and arguing that they had
discovered what they called ''dark energy'', which of course is another name for
creation.  Not surprisingly this result can be easily understood in the framework of the
cyclic universe model (cf Viswakarma et al 2005), Narikar et al (2002) in which there is
no big bang.

\underline{SUMMARY}

In conclusion we summarize the strengths and weaknesses of the hot big bang model, and
its alternatives.

(a)  The major arguments in favor of the hot big bang model are that, on the surface at
least, we can understand the expansion in terms of a Friedmann model, and if the
parameters are chosen correctly we can explain out the details of the CMB as it is found
at present.  The very small structure in the radiation can be explained provided that it
is supposed that the microwave field has interacted with the lumpy matter component early
in the expansion, and that the matter had density fluctuations of unknown (probably
quantum) origin.  Moreover we can explain the abundances of the light isotopes made in
the initial explosion.

However to explain the existence of galaxies, many unproven assumptions have to be made.

As is the case for all origin problems we must assume that there are initial density
fluctuations.  Early on it is necessary to invoke and inflationary phase for which we
have no basic theory though it is an attractive idea.  Also, in order to explain the
abundance of the light isotopes and the flatness problem and so on we must invoke the
presence of non-baryonic matter for which there is no direct evidence at all.

It is important to stress that this component must be dominant in forming galaxies, -
without it galaxy formation cannot occur.  Even then to make the large scale structure
agree with observations other parameters must be used, including ''biasing,'' and the
falling together of small galaxies to form larger ones.

It is also necessary to ignore any of the evidence which points to radical events taking
place in the nuclei of galaxies.  Redshifts everywhere must be cosmological in origin and
even effects like the Tifft effect, well-established by Gutherie and Napier (1996) and
his colleagues showing that there is a small non-cosmological component in the redshifts
or normal galaxies, must be ignored.

(b)  The major evidence for a cyclic universe model is first that observations show that
the major components of the universe are the galaxies which are condensed regions of
matter (lumps).  They are the primary energy sources, and all other constituents - other
photon fields, and diffuser matter, etc., arise in them and are ejected from them.

The fact that the energy density in the CMB is in such good agreement with the amount of
energy released in building the observed helium abundance through hydrogen burning (based
simply on estimates of the masses of the galaxies and the observed distance scale) shows
that the time scale for the process must be long and must be $>(H_{o}^{-1})$.  Thus the
observed expansion seen at this epoch does not have anything to do with the Friedmann
solution to Einstein's equations.  Instead the time scale leads directly to a cyclic
universe model, originally described as the quasi-steady state model (QSSC)(Hoyle, ET AL
2000; Narlikar and Burbidge, 2007) which has a characteristic cycle time of $\approx50
Gyr$ and an expansion timescale $\approx10^{12}-11^{13}$ years.  In the hot big bang
creation takes place once at t=0, but in the cyclic universe creation is taking place
continuously.

Rather that being synthesized in a big bang, the light isotopes are made following
creation in little big bangs in many galaxies over a long time scale.  We have shown
earlier that the helium can be made from hydrogen burning in galactic nuclei.  The
deuterium must be made in flares in stellar atmospheres.

The attraction of this aspect of the model is that it suggests that \underline{all} of
the elements are made in stars, and not just the heavier isotopes as was originally
proposed by Cameron and $B^{2}FH$ (Burbidge and Hoyle 1998).  There is one other
possibility for nucleosynthesis in galactic nuclei.  As was originally realized by
Wagoner, Fowler and Hoyle, and as described in Chapter 10 of Hoyle et al (2000), the
physical conditions close to the condensed centers where little big bangs can occur, are
$\rho\simeq10^{9}gm cm^{-3}$ and $T_{9}=10$.  With a timescale of $\approx1-^{-13}$ sec,
and a baryon/photon ration $\approx1$ ''local'' primordial nucleosynthesis reactions can
produce the observed abundances of the light isotopes.

In the cyclic universe we shall expect to see a build-up of dark baryonic matter, but
there is no reason to invoke the presence of dark non-baryonic matter.

The main difficulty with this model is that we have no detailed understanding of the
creation process, though the advantage here as compared with the big bang is that the
manifestations of creation in the form of active galaxies appear to be taking place all
around us at every epoch.

A second difficulty, believed by many, is that we do not have the same detailed
understanding of the energy distribution in the observed CMB as is claimed for the big
bang (cf Spergel et al 2006).  This is true, though some progress has been made (Narlikar
et al 2003), but we believe that this is because to obtain the agreement between theory
and observation in the conventional scheme many parameters which are really unknown, have
to be chosen, and in addition, perhaps overwhelmingly, because hundreds of cosmologists
are working on the big bang compared with a small handful who are looking at the
alternatives.

(c)  Other cosmological models are yet to be discovered.  If it is ultimately confirmed,
or even seriously believed, that Newton's law needs to be modified at large distances, as
is suggested in MOND, basic changes will have to be made to what is being proposed in (a)
and (b).  And if we are forced to develop new physics to account for anomalous redshifts
which I believe are real, this will undoubtedly tell us something fundamentally new about
the origin of mass.

\underline{CONCLUSION}

Cosmology is drive by observational discoveries, and after ninety years we are still a
long way from a real understanding of the large scale structure of the universe.  I
believe that a major step backward was made in the 1960s when the leaders in the field
decided that the basic problem had been solved, thus heavily biasing attitudes, outlook
and funding in one direction only.

This meeting is dedicated to the memory of Denis Sciama.  I knew him well in Cambridge
and it is very good that we remember him here.  Denis made major contributions,
particularly in his encouragement and training of some of the leaders today.  But to me
Denis was unique in one way since he was the only leading cosmologist I have ever met who
changed his mind$\!$  Originally he passionately believed in the steady-state but later
he was equally eloquent about the big bang.

Will such an event ever happen again?

\end{document}